\documentclass{article}%
\usepackage{amsmath}
\usepackage{amsfonts}
\usepackage{amssymb}
\usepackage{graphicx}
\usepackage{hyperref}%
\providecommand{\U}[1]{\protect\rule{.1in}{.1in}}
\begin{document}
\vskip 2cm
\centerline
{\bf The role of the 1.5 order formalism and the gauging of  spacetime groups   }
\centerline {\bf  in  the development of   gravity and supergravity theories}

%{\bf A brief review  of the 1.5 order "formalism",  formulating gravity and supergravity  }
%\centerline {\bf as  gauge theories  and the most general coupling of  simple supergravity to matter }
\vskip 1cm
\centerline{\bf Ali H. Chamseddine$^{1}$\  and  Peter West$^{2}$ }
 \vskip 1.2cm
	\centerline{$^{1}$ {\it Physics Department, American University of Beirut, Lebanon}}
	\vskip0.4cm
    	\centerline{$^{2}$ {\it Department of Mathematics, King's College, London WC2R 2LS, UK}}
			
%\end{center} 
\vskip 1.7cm
%\centerline{and}
%\vskip 0.5cm
%\centerline{Peter West,}
%\centerline {,}
%\centerline{}
\leftline{\sl Abstract}
The 1.5  formalism   played a key role in the discovery of supergravity and it has been   used to prove the invariance of essentially all supergravity theories under local supersymmetry. It emerged  from  the gauging of the super Poincare group   to find supergravity. We review both of these developments as well as the auxiliary fields for simple supergravity  and its most general coupling to matter using the tensor calculus. 
\noindent
\vskip 6cm
emails: chams@aub.edu.lb, peter.west540@gmail.com
\vfill

\eject

\medskip
A theory of supergravity was first proposed by Ferrara, Freedman and van Nieuwenhuizen in a paper [1] entitled "Progress towards a theory of supergravity" which contained the vierbein $e_\mu{}^a$ and the gravitino $\psi_\mu{}_\alpha$. They proposed the action 
$$
A=\int d^4x\left\{{e\over2\kappa^2}R-{1\over2}\bar\psi_\mu 
R^\mu \right\}\eqno(1)$$
and the local supersymmetry transformations 
$$
\delta e^{\ a}_\mu=\kappa\bar\varepsilon\gamma^a\psi_\mu ,\ \  
\delta\psi_\mu=2\kappa^{-1}D_\mu\big(w(e,\psi)\big)\varepsilon
\eqno(2)$$
In these equations 
$$
R=R_{\mu\nu}^{\ \ ab}e_a^{\ \mu}e_b^{\ \nu}, \ \ R^\mu=\varepsilon^{\mu\nu\rho\kappa}i\gamma_5\gamma_\nu
D_\rho\big(w(e,\psi)\big)\psi_\kappa , \ \ R_{\mu\nu}^{\ \ ab}{\sigma_{ab}\over 4}=[D_\mu,D_\nu] ,\ \gamma_\mu= e_\mu{}^a\gamma_a 
\eqno(3)$$
where the Lorentz covariant derivative is given by 
$$
 D_\mu\big(w(e,\psi)\big)=\partial_\mu+w_{\mu ab}{\sigma^{ab}\over 4} ,
\eqno(4)$$ 
with   
$$
w_{\mu ab}={1\over2}e^\nu_{\ a}(\partial_\mu e_{b\nu}-\partial_\nu
e_{b\mu})-{1\over2}e_b^{\ \nu}(\partial_\mu e_{a\nu}-\partial_\nu
e_{a\mu})-{1\over2}e_a^{\ \rho}e_b^{\ \sigma}(\partial_\rho e_{\sigma
c}-\partial_\sigma a_{\rho c})e_\mu^{\ c}
$$
$$
\quad+{\kappa^2\over4}(\bar\psi_\mu\gamma_a\psi_b
+\bar\psi_a\gamma_\mu\psi_b-\bar\psi_\mu\gamma_b\psi_a)
\eqno(5)$$
They showed that the action was invariant up to, and including, cubic terms  in the gravitino and stated that the quintic terms vanished using a computer programme. They also showed that the supersymmetry transformations closed 
up to terms cubic  in the gravitino if one uses the equations of motion   This theory was in second order formalism as it contained the vierbvein and gravitino but not the spin connection as an independent field. This supergravity is often referred to as $N=1$, $D=4$ supergravity, or simple supergravity. 
\par
A bit later, a theory of supergravity involving the vierbein $e_\mu{}^a$,  the gravitino $\psi_\mu{}_\alpha$ and a spin connection $\omega_\mu{}^{ab}$ was proposed [2]. This paper proposed an action which was in first order formalism, that is, the spin connection was an independent field and had an independent  supersymmetry transformation. These authors showed that their  theory  did not have anomalous characteristics of its surfaces of propagation. In other words  it has a consistent propagation.  It was  known that the propagation of a spin 3/2 particle coupled to a spin 1 particle was not consistent and the same was suspected to be the case for generic higher spin theories. Reference [2]  contains a two sentence discussion of the invariance of the action under the supersymmetry transformations that  uses equation (11) which is, in effect,  an  equation of motion. The paper also does not discuss the closure of the supersymmetry transformations.   There has subsequently been almost no work on supergravity in first order formalism and it remains an interesting open problem to develop it further. 
\par
A different approach to supergravity was taken some months  later in the paper [3]. This paper considered the gauge theory of the super Poincare group,  which has the generators $P_a$, $Q_\alpha$ and $J_{ab}$ and the algebra 
$$
\left[  P_{a},P_{b}\right]     =J_{ab} ,\ \ 
\left[  P_{a},J_{bc}\right]     =\left(  \eta_{ab}P_{c}-\eta_{ac}
P_{b}\right)  
$$
$$
\left[  J_{ab},J_{cd}\right]     =\left(  \eta_{ad}J_{bc}-\eta_{ac}
J_{bd}-\eta_{bd}J_{ac}+\eta_{bc}J_{ad}\right) ,
$$
$$
\left\{  Q_{\alpha},Q_{\beta}\right\}     =-2\left(  \gamma^{a}C^{-1}\right)
_{\alpha\beta}P_{a} ,\ \ 
\left[  J_{ab},Q_{\alpha}\right]     =-{1\over 2}\left(  \gamma
_{ab}Q\right)  _{\alpha} , \ \ \left[  P_{a},Q_{\alpha}\right]   =0
\eqno(6)$$

 As such  they introduced the connection 
$$
A_\mu= e_\mu{}^a P_a -{1\over 2} \omega_\mu{}^{ab}J_{ab}+{1\over 2} \bar \psi^\alpha Q_\alpha
\eqno(7)$$
and the corresponding field strengths defined by $ [\hat D_\mu,\hat D_\nu ]=-R_{\mu\nu}{}^a
P_a +{1\over 2}R_{\mu\nu}{}^{ab} J_{ab} +{1\over 2}\bar \Psi_{\mu\nu} Q$ where $D_\mu =\partial_\mu -A_\mu$. 
The field strengths are  
$$
R_{\mu\nu}{}^a= \partial_\mu e_\nu^a- \partial_\nu e_\mu^a+
\omega_{\mu}{}^a{}_c e_\nu^c -\omega_{\nu}{}^a{}_c e_\mu^c
+{1\over 2}\bar\psi_\mu\gamma^a\psi_\nu,
\quad R_{\mu\nu}{}^{a b}=\partial _\mu \omega_\nu{}^{ab}
+\omega_\mu{}^{ac}\omega_\nu{}_c{}^{b}- (\mu\leftrightarrow \nu),
$$
$$
\Psi_{\mu\nu}= (\partial _\mu
-{1\over 4}\gamma_{cd}\omega_\mu{}^{cd})\psi_\nu
-(\mu\leftrightarrow \nu)\equiv D_\mu \psi_\nu -(\mu\leftrightarrow \nu)
\eqno(8)$$
The variations of the fields under gauge transformations of the form 
$\Lambda = v^a P_a -{1\over 2}  \omega^{ab}J_{ab}+ {1\over 2}\bar \epsilon ^\alpha Q_\alpha$ 
are given by $\delta A_\mu= \partial_\mu \Lambda -[A_\mu, \Lambda ]$ and so the individual fields transform as 
$$
\delta e_\mu^a= \partial_\mu v^a- \omega^{a}{}_c e_\mu ^c
+\omega_\mu{}^{ac} v _c+{1\over 2}\bar \epsilon\gamma^a\psi_\mu,\quad
\delta \omega_\mu{}^{ab} =\partial_\mu \omega_{}^{ab}
-(\omega^{ac} \omega_\mu{}_c{}^{b}-\omega^{bc} \omega_\mu{}_c{}^{a}),\quad
$$
$$
\delta \psi_\mu =2(\partial _\mu
-{1\over 4}\gamma_{cd}\omega_\mu{}^{cd})\epsilon 
+{1\over 4}\gamma_{cd}\omega^{cd}\psi_\mu\equiv  D_\mu \epsilon  
+{1\over 4}\gamma_{cd}\omega^{cd}\psi_\mu
\eqno(9)$$
\par
In reference [3] the action was taken to be linear in the field strengths and the unique such action which is invariant under local Lorentz transformations is of the form  
$$- {1\over 8} \int d^4 x\ \epsilon ^{\mu\nu\rho\lambda }(\epsilon_{abcd
}e_\mu^a e_\nu^b R_{\rho\lambda}{}^{cd} - 2if\bar \psi_\mu
\gamma_5\gamma_\nu
\Psi_{\rho\lambda} ).
\eqno(10)$$
where $f$ is a constant. Since the action is not of the form of the squares of the field strengths it can not be invariant under the above gauge transformations. However, the authors of reference [3]   only demanded invariance up to the  the condition
$$
R_{\mu\nu}^a=0
\eqno(11)$$ 
\par
We will now vary  the action of equation (10) under the transformations of equation (9)  subject to the  condition of equation (11). The argument follows the steps of reference [3] except that, for simplicity,  we will take  $f=-1$, which is the value determined from the variation. The variation of the Einstein part is given by
$$
\delta \int {e\over 2 \kappa^2} \big(e_a{^\mu} e_b{^\nu} R_{\mu\nu}{^{ab}}\big)\,d^4
x  = \int d^4 x\!\left\{{1\over \kappa} \{\bar
{\varepsilon}\gamma^\mu\psi_a
\}\bigg\{-\!R_\mu{^a}+{1\over 2}e_\mu{^a} R\bigg\}\!\right\}
\eqno(12)$$
while the variations of the Rarita-Schwinger part of the action gives the following three terms
$$
\delta \int
\left(-{i\over 2}\bar{\psi}_\mu\gamma_5 e_\nu{}^a\gamma_a D_\rho
\psi_\kappa\varepsilon^{\mu\nu\rho\kappa}\right)\,d^4 x = \int d^4
x\bigg\{-{i\over \kappa}\bar{\varepsilon}
\overleftarrow{D}_\mu\gamma_5\gamma_\nu D_\rho\psi_\kappa
\varepsilon^{\mu\nu\rho\kappa} \bigg\}
$$
$$
\bigg\{-{i\over \kappa}\bar{\psi}_\mu\gamma_5\gamma_\nu\overrightarrow{D}_\rho
D_\kappa \varepsilon
\varepsilon^{\mu\nu\rho\kappa}-{\kappa\over 2}
i\bar{\varepsilon}\gamma^a \psi_\nu\bar{\psi}_\mu \gamma_5
\gamma_a D_\rho \psi_\kappa
\varepsilon^{\mu\nu\rho\kappa}\bigg\}
\eqno(13)$$
Flipping the spinors using their Majorana property we find that the second term of the above equation takes the form
$$
-{i\over 8 \kappa}\bar{\psi}_\mu \gamma_5
\gamma_\nu R_{\rho\kappa}{}^{cd}
\sigma_{cd}\varepsilon\varepsilon^{\mu\nu\rho\kappa} =
-{i\over 8 \kappa}
\bar{\varepsilon}\sigma_{cd}\gamma_\nu\gamma_5 \psi_\mu
R_{\rho\kappa}{^{cd}}\varepsilon^{\mu\nu\rho\kappa}
\eqno(14)$$
Integrating the first term of Eq. (13) by parts and neglecting surface terms we find that it is given by 
$$
 {i\over \kappa}\bar{\varepsilon}\gamma_5 [D_\mu,
\gamma_\nu]D_\rho\psi_{\kappa}\varepsilon^{\mu\nu\rho\kappa}
+{i\over \kappa}\bar{\varepsilon}\gamma_5\gamma_\nu D_\mu
D_\rho\psi_\kappa\varepsilon^{\mu\nu\rho\kappa}
\eqno(15)$$
Using equation (3) the second of these terms is given by 
$$
{i\over 8 \kappa}\bar{\varepsilon}\gamma_5\gamma_\nu
R_{\rho\kappa}{^{cd}} \sigma_{cd}\psi_\mu
\varepsilon^{\mu\nu\rho\kappa}
\eqno(16)$$

The term  given in equation (16) and that in equation (14) add together to give the result
$$
+{i\over 2\cdot 4 \kappa}\bar{\varepsilon}\gamma_5(\gamma_\nu
\sigma_{cd} + \sigma_{cd}\gamma_\nu)\psi_\mu
R_{\rho\kappa}{^{cd}}\varepsilon^{\mu\nu\rho\kappa}
={1\over 4 \kappa}
\bar{\varepsilon}\gamma_f\psi_\mu\varepsilon_{f\nu c
  d}\varepsilon^{\mu\nu\rho\kappa}R_{\rho\kappa}{}^{cd}
$$
$$
=-{1\over 2\kappa}\bar{\varepsilon}\gamma^a\psi_\mu\big\{e_a{^\mu}
R - 2 R_a{^\mu}\big\} e
\eqno(17)$$
which  exactly cancels the variation of the Einstein~action given in Eq. (12).

Consequently, we are just left with the first term of equation  (15) and the last term of equation  (13), Performing a Fierz transformation (see, for example,  the appendix of reference [5] for details) on the latter term it becomes
$$
-{\kappa\over 2\cdot 4}i\bar{\varepsilon}\gamma^a \gamma_R
\gamma_a \gamma_5 D_\rho
\psi_\kappa\varepsilon^{\mu\nu\rho\kappa}\bar{\psi}_\mu\gamma_R\psi_\nu = +{\kappa\over 4}
i\bar{\varepsilon}\gamma_c\gamma_5D_\rho\psi_\kappa\varepsilon^{\mu\nu\rho\kappa}\bar{\psi}_\mu\gamma^c\psi_\nu
\eqno(18)$$
The first term in  equation (15) is most easily  evaluated by going to inertial coordinates,
that is, we set $\partial_\mu e_\nu{^a} = 0$; it becomes
$$
{i\over 4 \kappa} \bar{\varepsilon}\gamma_5
[\sigma^{cd},\gamma_\nu]
w_{\mu cd}D_\rho\psi_\kappa\varepsilon^{\mu\nu\rho\kappa}={i\over \kappa}\bar{\varepsilon}\gamma_5\gamma^c D_\rho\psi_\kappa w_{\mu c\nu} \varepsilon^{\mu\nu\rho\kappa}
$$
$$
={\kappa\over 4}i\bar{\varepsilon}\gamma_5 \gamma^c D_{\rho}
  \psi_{\kappa} \bar{\psi}_\mu \gamma_c \psi_\nu
  \varepsilon^{\mu\nu\rho\kappa}
  \eqno(19)$$
  This term cancels with that of equation  (18). This completes the proof of invariance.  
 \par
 Adopting the constraint of equation (11) was somewhat unconventional and we now discuss it in more detail. 
 Equation (11) allows one to express the spin connection in terms of  the viebein and gravitino, indeed this is all the information it contains. The resulting expression is  nothing but the equation of motion of the spin connection of the action of equation (10) with $f=-1$. As such  the constraint of equation (11) takes the theory from  the  first to second oder formalism. Indeed adopting  this value for  the spin connection, equation (11) is  identically true. Enforcing   the condition of equation (11),  the action of equation (10)  and the transformations of the veirbein and gravitino of equation (9) are just those found in reference [1]. Thus gauging the super Poincare group leads to the supergravity theory discovered in reference [1], that is, the same action and transformation laws, however, it had the great advantage that it also showed that it was invariant under the local supersymmetry transformations. 
 \par 
It remains to comment on the fact that in the steps above we essentially did not vary the spin connection in the action. 
 Varying the action of equation (1),  or equivalently equation (10) with $f=-1$, we have 
$$
\delta A= \int d^4x ({\delta A\over \delta e_\mu{}^a} \delta e_\mu{}^a+ {\delta A\over \delta \psi_\mu{}_\alpha}\delta \psi_\mu{}_\alpha+ {\delta A\over \delta \omega_\mu{}^{ab}}\delta \omega_\mu{}^{ab})
\eqno(20)$$
Since we are in second order formalism, the variation of $ \omega_\mu{}{}^{ab}$ is just that found by varying the vierbein and graviton upon which it depends. Just to be  completely clear 
$$
\delta \omega_\mu{}^{ab}= {\delta \omega_\mu{}^{ab}\over \delta e_\mu{}^c}\delta e_\mu{}^c+ 
 {\delta \omega_\mu{}^{ab}\over \delta \psi_\mu{}_\alpha}\delta \psi_\mu{}_\alpha
\eqno(21)$$
where the variation of the vierbein and graviton are those of equation (2).  However, the last term of the variation of the action,  given in equation (20),  vanishes. as 
$$
 {\delta A\over \delta \omega_\mu{}^{ab}}={e\over 2}R_{\kappa\lambda}^{\quad
c}\left(  e_{c}^{\lambda}\left(  e_{a}^{\mu}e_{b}^{\kappa}-e_{b}^{\mu}%
e_{a}^{\kappa}\right)  +e_{c}^{\mu}e_{a}^{\kappa}e_{b}^{\lambda}\right)=0
 \eqno(22)$$
as a consequence of the constraint equation (11). Thus in effect one  does not have to vary the spin connection in the action. In second order formalism  $\omega_\mu{}^{ab}$ is not an independent field but is given in terms of  the vierbein and graviton. Indeed, it  is the equation of motion of $\omega_\mu{}^{ab}$ that determines the spin connection in this way. As such equation (22) is not an equation of motion but an identity. A straight forward account of the   invariance  of  the action was  given in reference [6]. This paper adopted the steps in reference  [3], that is  equations (12-19),  but also implemented   equation (22). 
\par
With the above  steps  the discovery of supergravity was complete, the  transformations rules of paper [1] were shown to be an invariance  of the action of the seminal reference  [1] using the usual analytic methods given in reference  [3]. The advantage of this was that any reader could  verify that the action was invariant so opening up the way to further discoveries. The method of paper [3] has been used to show the invariance of all supergravity actions in all dimensions. 
\par
 At some point the  above procedure was given the name the 1.5 order formalism, a name by which it  is now known. However, the supergravity of  theory of reference [3] and indeed the discussion of equations (12-19) is  in second order formalism as we have  implemented the constraint of equation (11). The proof of invariance presented in reference [3]  is really a method and not a  formalism. This aspect has mislead some authors,  such as in reference [7]. 
\par
The method of gauging the supersymmetry algebra to derive supergravity 
presented in reference [3] was simultaneously also  presented in an alternative form
based on gauging the Orthosymplectic algebra $OSP\left(  1,4\right)  $ which has  the generators $P_{a},$ $J_{ab}$ and $Q_{\alpha}$ [7]. In this work it was shown how to
obtain the $N=1$ supergravity fields $e_{\mu}^{a},$ $\omega_{\mu}^{ab}$ and
$\psi_{\mu}$  as gauge fields of $OSP\left(  1,4\right)  $. The transformations of the fields were calculated and it was shown  how to  recover the above results based on the super Poincare algebra of reference [3] by rescaling $P_{a}\rightarrow RP_{a},$ $Q_{\alpha}\rightarrow\sqrt
{R}Q_{\alpha},$ $J_{ab}\rightarrow J_{ab}$ then taking the infinite radius
limit $R\rightarrow\infty$.   This  was taken up in the work  of MacDowell and Mansouri  who completed the calculation using the  Orthosymplectic algebra and constructed an action based on field strengths squared [8]. Despite the aesthetic appearance of the action one still has to  impose equation (11).  A similar calculation was also presented in reference [9]. 
\par
A restriction of the results of reference [3]  can also be used to find Einstein's theory of general relativity in a very simple way. We begin by gauging  just the Poncar\' e group and so set to zero the gravitino in the above equations. The resulting theory 
has just local translations and rotations. We also  adopt the constraint of equation (11) with the gravitino set to zero. 
 The remarks above about the spin connection and the proof of invariance apply in the same way to the case of pure gravity. A review of the above gauging of the Poincare group can be found in section 13.1.3  in the book  of reference [4]

 \par
Adopting the condition of equation (11) is rather unconventional as it breaks by hand  the gauge symmetry and in particular the local translations. However, one can write the  gauge transformation of the vierbein of equation (8) as 
$$
\delta e_\mu{}^a= \partial_\mu \xi^\lambda e_\lambda{}^a + \xi^\lambda \partial_\lambda e_\mu{}^a 
-(\xi^\lambda w_\lambda {}^a{}_b) e_\mu{}^b - {1\over 2} (\xi^\lambda \psi_\lambda ) \gamma^a \psi_\mu+ \xi^\lambda R_{\mu\lambda}
\eqno(23)$$
where $\xi^\mu= e^\mu{}_b v^b$. We  recognise this transformation as a diffeomorphism, a local Lorentz transformation  and a local supersymmetry transformation provided the constraint of equation (11) holds. Thus we have the paradoxical result that imposing the condition of equation (11) we find that the local translations become  a combination of a diffeomorphism,  a local Lorentz  and a local supersymmetry transformation which are symmetries of the final theory.  This feature appears in the other applications of  reference [3] to the   gauging other spacetime groups. A way to proceed without taking the constraint of equation (11),  and so not breaking the gauge symmetry by hand,  was to introduce some more fields which are constrainted [14]. 
\par
It will be instructive to recall previous developments on the connection between the Poincare group and general relativity. The viebein was introduced into general relativity by Herman Weyl in 1929 [15].  In the papers [16],  [17] and  [18 ] it was  shown that the spin connection of general relativity in first order form could be thought of as the gauge field for the Lorentz group, indeed the Riemann curvature was just the corresponding field strength. The  authors of references  [17] and  [18] also considered what they called the gauge theory of the Poincare group. In this approach they  took the well known coordinate transformations of the Poincar\'e group on Minkowski spacetime 
$$
x^\mu\to x^\mu + \omega^\mu{}_\nu x^\nu+ a^\mu
\eqno(24)$$
and let the constant parameters $ \omega^\mu{}_\nu $ and $a^\mu$ be local, that is,  depend on spacetime. As they pointed out,  in this way one introduces a diffeomorphism. This approach has been extensively pursued and there is now a substantial literature, see for example [19]. This literature  is not the same as taking  the gauge theory of the Poincare group  in the sense of Yang-Mills which was the approach of reference [3]. As is well known the unique action of  a Yang-Mills theory consists of its  field strength squared and this is not the case of gravity. This fact had, perhaps,  put off researchers from carrying out a direct gauging of the Poincare, In this context it is amusing to read the first sentence of the introduction  of reference [3]; the authors were just PhD students! 
\par
The advantage of the gauge approach of reference [3] was that it  provided   a simple way to construct  Einstein's theory of general relativity if one gauge the Poincare group, and supergravity if one gauged the super Poincare group. Indeed this gauging approach was used to construct  the theories of super conformal gravity [20], the  super conformal tensor calculus as well as  gravity and supergravity  in three dimensions [30], ...   It also  underlies the construction of higher spin theories [21] where one gauges  an infinite dimensional gauge group rather than the Poincare, or super Poincare group. 
\par
One draw back of the original formulation of supergravity [1,3] was that the local supersymmetry algebra only closed when one used the equations of motion. This meant that the coupling of supergravity to any super matter, that is, any combination of   the super Yang-Mills and Wess-Zumino models,  was a formidable task. Indeed,  the task had to be repeated for each new matter model as the equations of motion were different and, as a consequence, so were  the local supersymmetry transformations. This changed with the discovery of the auxiliary fields $M$, $N$ and $b_\mu$ for the simplest supergravity in four dimensions [10,11]. 
The action was given by 
$$A=\int d^4x\left\{{e\over2\kappa^2}R-{1\over2}\bar\psi_\mu 
R^\mu-{1\over3}e(M^2+N^2-b_\mu b^\mu)\right\}
\eqno(25)$$
and the transformations by 

%$$\eqalignno{
\begin{flalign*}
\delta e^{\ a}_\mu&=\kappa\bar\varepsilon\gamma^a\psi_\mu\cr
\delta\psi_\mu&=2\kappa^{-1}D_\mu\big(w(e,\psi)\big)\varepsilon+i\gamma_5\left(b_\mu-{1\over3}\gamma_\mu/\!\!\!b\right)\varepsilon
-{1\over3}\gamma_\mu(M+i\gamma_5N)\varepsilon\cr
\delta M&=-{1\over2}e^{-1}\bar\varepsilon\gamma_\mu
R^\mu-{\kappa\over2}i\bar\varepsilon\gamma_5\psi_\nu
b^\nu-\kappa\bar\varepsilon
\gamma^\nu\psi_\nu
M+{\kappa\over2}\bar\varepsilon(M+i\gamma_5N)\gamma^\mu\psi_\mu\cr
\delta N&=-{e^{-1}\over2}i\bar\varepsilon\gamma_5\gamma_\mu
R^\mu+{\kappa\over2}\bar\varepsilon\psi_\nu b^\nu-\kappa\bar\varepsilon
\gamma^\nu\psi_\nu
N-{\kappa\over2}i\bar\varepsilon\gamma_5(M+i\gamma_5N)
\gamma^\mu\psi_\mu\cr
\delta
b_\mu&={3i\over2}e^{-1}\bar\varepsilon\gamma_5\left(g_{\mu\nu}-{1\over3}\gamma_\mu\gamma_\nu\right)R^\nu+\kappa\bar\varepsilon
\gamma^\nu
b_\nu\psi_\mu-{\kappa\over2}\bar\varepsilon\gamma^\nu\psi_\nu b_\mu\cr
&\quad-{\kappa\over2}i\bar\psi_\mu\gamma_5(M+i\gamma_5N)
\varepsilon-{i\kappa\over4}\varepsilon_\mu^{\
bcd}b_b\bar\varepsilon\gamma_5
\gamma_c\psi_d&(26)
\end{flalign*}
%$$
These transformations  closed without the use of equations of motion, namely 
$$
[\delta_{\varepsilon_1},\delta_{\varepsilon_2}] =\delta_{\rm
supersymmetry}(-\kappa\xi^\nu\psi_\nu)+\delta_{\rm general\
coordinate}(2\xi_\mu)
$$
$$
+\delta_{\rm Local\
Lorentz}\left(-{2\kappa\over3}\varepsilon_{ab\lambda\rho}
b^\lambda\xi^\rho -{2\kappa\over3}\bar\varepsilon_2\sigma_{ab}
(M+i\gamma_5N)\varepsilon_1+2\xi^dw_d^{\ 
ab}\right)
\eqno(27)$$
where $\xi_\mu=\bar\varepsilon_2\gamma_\mu\varepsilon_1$. 

It was straightforward to extend the proof of invariance of the action without auxiliary fields [3] to include them. 
 With this last step we now have a supergravity theory which possess local supersymmetry transformations that satisfy a closing algebra which is independent of any specific dynamics and leave the action invariant. Of course this is  the usual situation with symmetries before that of supersymmetry. It was   straightforward to quantise the simple supergravity  theory using the usual BRST techniques [12].  This  contrasts with the statements  given in  reference [13] which finds not only this result,  but the 1.5 formalism itself to be troublesome. 
\par
The discovery of the auxiliary fields allowed the construction of a tensor calculus for supergravity  which made it easy to compute the most general coupling of $D=4,$ $N=1$ supergravity to the most general matter, which,   in turn,   paved the way to construct a realistic spontaneously broken supersymmetric model. We will now explain how the tensor calculus was constructed [22,23].   Matter consists of chiral multiplets (Wess-Zumino) $\Sigma^a$ and vector multiplets $V$. The 
  chiral  multiplets have the field content 
$$
\Sigma^{a}=\left(  z^{a},\chi_{L}^{a},h^{a}\right)
\eqno(28)$$
where $z^{a}=A^{a}+iB^{a}$ are complex scalar fields, $\chi_{L}^{a}$ are
left-handed Weyl spinors and $h^{a}=F^{a}+iG^{a}$ are complex auxiliary
fields. Taking the complex conjugate of the above chiral super multiplet,  we find it contains a  spin zero field $z_a$, which is the complex conjugate of $z^a$,  and also a  spinor  of the opposite chirality. The index $a$ on $\Sigma^{a}$ is an internal symmetry index which corresponds to fact that   $\Sigma^{a}$  can belong to a  representation of a gauge group $G$. 
\par
The vector multiplet $V$ is  real and has the components 
$$
V=\left(  C,\zeta,H,K,v_{\mu},\lambda,D\right)  
\eqno(29)$$
whcih  belong to  the adjoint representation of the gauge group $G.$ In this equation 
$\zeta,$ $\lambda$ are Majorana spinors and $C,$ $H,$ $K$ are scalars while $D$, which is also a 
scalar,  is an auxiliary field. These super multiplets have been    used to construct realistic models of nature that have rigid supersymmetry. The quarks, leptons and Higgs are expected to be contained in chiral super multiplets   while the vector super multiplets contain the  spin one gauge particles. It is far from clear how to break supersymmetry in the context of rigid supersymmetry. 
\par
As we have mentioned the introduction of the auxiliary fields lead to a theory of simple supergravity whose fields possessed  transformations  that closed without the use of the equations of motion. This  is the supergravity analogue of the closure of two general coordinate transformations in general relativity. With this result the  chiral and vector multiplets of rigid supersymmetry could then  be generalised to be multiplets of this local supersymmetry, that is carry a representation of this local algebra. In particular,    their supersymmetry transformations should have a local spinor parameter and they must have   a closing algebra that is the same as that for the supergravity fields, in other words that of  equation (26). To achieve this  their transformations must be   extended to include terms involving the supergravity fields. Given these local chiral and vector  super multiplets we can, as in general relativity, construct a tensor calculus. In other words we can construct local chiral and vector multiplets out of products of such multiplets. The precise formulae can be found in references [22] and [23],  or the review of chapter thirteen  in the book of reference [5].  
\par
The final step in the construction of the tensor calculus is to construct the supersymmetric invariants for the chiral and vector multiplets.  For rigid supersymmetry these are just given by the integrals over spacetime  of the two auxiliary fields $F$ and $D$ respectively. These are called the $F$ and $D$ terms. Their  generalisation to be invariant under local supersymmetry are easy to find given the local supersymmetry transformations of the fields [22], [23]. The invariant for the chiral super multiplet, the $F$ term,  is given by   
$$
A_{F}=\int d^4 x  e \left(  F-\left(  MA+NB\right)  +{1\over 2}\overline{\psi}_{\mu
}\gamma^{\mu}\chi+{1\over 4}\overline{\psi}_{\mu}\gamma^{\mu\nu}\left(
A+i\gamma_{5}B\right)  \psi_{\nu}\right) 
\eqno(30)$$
While the invariant for the vector super multiplet, the $D$ term,   is given by 
$$
A_{D}  =\int d^4 x  e\{D-{i\kappa\over 2}\overline{\psi}_{\mu}\gamma^{\mu}\gamma
_{5}\lambda+{2\over 3}\left(  MK-NH\right)  -{2\kappa\over 3}A_{\mu}\left(
b^{\mu}+{3\kappa e^{-1}\over 8}\epsilon^{\mu\nu\rho\sigma}\overline{\psi
}_{\nu}\gamma_{\rho}\psi_{\sigma}\right) 
 $$
 $$
 -{\kappa\over 3}\overline{\zeta}\left(  i\gamma_{5}\gamma_{\mu}R^{\mu
}+{3\kappa\over 8}\epsilon^{\mu\nu\rho\sigma}\psi_{\mu}\overline{\psi}_{\nu
}\gamma_{\rho}\psi_{\sigma}\right)  -{2\kappa^{2}\over 3}e^{-1}L_{SG}\}
\eqno(31)$$
where $L_{SG}$ is the Lagrangian of simple supergravity which can be read off from equation (25). We observe  that all the  fields of the relevant  super multiplet occur in these local   $F$ and $D$ terms, as do all the supergravity fields.  A complete discussion of the tensor calculus can be found in chapter thirteen  of  the book of reference [5]. 
\par
Using the tensor calculus it is easy to find the most general coupling of super matter to simple supergravity; one just has to apply the formulae for the composition of the super multiplets and the above density formulae of equation (30) and (31). The resulting action  can be expressed in terms of three functions $g\left(  z^{a}\right)  ,$ $\phi\left(  z^{a}
,z_{a}\right)  $ and $\ f_{\alpha\beta}\left(  z^{a}\right)  .$ The
super potential function $g\left(  z^{a}\right)  $ is the lowest element of the
most general gauge singlet chiral multiplet formed out of the chiral multiplets
$\Sigma^{a}$.  We can write it as 
$$
g\left(  z^{a}\right)  =
A_{a_{1}a_{2}\cdots a_{m}}z^{a_{1}}z^{a_{2}}\cdots z^{a_{m}}%
\eqno(32)$$
The function $\phi\left(  z^{a},z_{a}\right)  $ represents the most general
gauge singlet vector  multiplet formed out of the chiral multiplet $\Sigma^{a}$
and its hermitian conjugate $\Sigma_{a}$ whose first components are $z^a$ and its complex conjugate $z_a$ respectively. It can be written as 
$$
\phi\left(  z^{a},z_{a}\right)  =B_{a_{1}\cdots a_{m}}^{\qquad b_{1}\cdots b_{n}}z^{a_{1}}\cdots
z^{a_{m}}z_{b_{1}}\cdots z_{b_{n}}%
\eqno(33)$$
The coefficients $A_{a_{1}a_{2}\cdots a_{m}}$ and $B_{a_{1}\cdots a_{m}%
}^{\qquad b_{1}\cdots b_{n}}$ are arbitrary parameters except that they
are chosen to maintain invariance under the gauge group  $G$. The function $f_{\alpha\beta
}\left(  z^{a}\right)  $ is the lowest component of a chiral function
transforming as the symmetric product of the adjoint representation of $G.$ In
most models  $f_{\alpha\beta}\left(  z^{a}\right)  $ is taken to be
$\delta_{\alpha\beta}.$ 
\par
The final action  resulting from the tensor calculus will contain the auxiliary fields for supergravity and matter multiplets. 
They can be eliminated using their equations of motion, although once this step is taken the resultant
Lagrangian will be invariant under supersymmetry transformation only after
using the equations of motion. The full Lagrangian is too long to list here,
so we will only write the bosonic part which can be transformed to be of  the form
[24], [25] [26]
$$
\int d^4 x e\{{1\over 2\kappa^{2}}R-{1\over 4}F_{\mu\nu}^{\alpha}F^{\mu
\nu\alpha}-{1\over \kappa^{2}}{\cal{G}},_{a}{}^{b}D_{\mu}z^{a}D^{\mu}
z_{b}
$$
$$
 -{1\over \kappa^{4}}e^{-{\cal{G}}}\left(  3+\left(  {\cal{G}}
^{-1}\right)  _{a}{}^{b}{\cal{G}}_{,}{}^{a}{\cal{G}},_{b}\right)  -
{1\over 8\kappa^{4}}\left\vert g_{\alpha}{\cal{G}}_{,a}\left(  T^{\alpha
}z\right)  ^{a}\right\vert ^{2}\}
\eqno(34)$$
where $F_{\mu\nu}^{\alpha}$ is the Yang-Mills field strength and the function
${\cal{G}}$ is defined by
$$
{\cal{G}}=3\ln\left(  -{\kappa^{2}\over 3}\phi\left(  z^{a},z_{a}\right)
\right)  -\ln\left(  {\kappa^{6}\over 4}\left\vert g\left(  z^{a}\right)
\right\vert ^{2}\right)
\eqno(35)$$
We have defined $D_{\mu}z^{a}$ as covariant derivative with respect to gauge
group $G,$ ${\cal{G}}_{,}^{a}={\partial{\cal{G}}\over \partial z_{a}},$
${\cal{G}},_{a}={\partial{\cal{G}}\over \partial z^{a}},$, 
${\cal{G}},_{a}{}^b={\partial^2{\cal{G}}\over \partial z^{a}\partial z_{b}} $,
$T^{\alpha}$  and the $g_{\alpha}$ are the matrices and gauge couplings associated with the
representation carried by $z^{a}$. 
\par
Unlike the case for rigid supersymmetry, the potential in equation (34) is no longer positive
definite because of the  corrections from supergravity. This fact is already apparent from the way the auxiliary fields occur in 
the supergravity action of equation (26) and the tensor calculus density formulae of equations (30) and (31). 
Realistic supergravity models can be constructed by considering a set of fields
$z^{A}=\left(  z,z^{a}\right)  $ where $z$ is a field belonging to the
super-Higgs sector, which is  the sector  responsible for supersymmetry breaking,  and
$z^{a}$ are the remaining matter fields. This can be achieved by considering
the super potential [24][28]
$$
g\left(  z^{A}\right)  =g_{1}\left(  z^{a}\right)  +g_{2}\left(  z\right)
\eqno(36)$$
where in the limit $\kappa\rightarrow0,$   there is no interaction between the fields $z^{a}$
and $z$. 
\par
For non-zero $\kappa$, these fields do have super-gravitational  interactions. The most dramatic
effect occurs in the $z^{a}$ sector, which  due to influence of the field $z$ , has soft breaking terms due to the super-Higgs effect. The superpotential $g_{2}\left(  z\right)  $ is taken to be of the form [28]
$$
g_{2}\left(  z\right)  =\kappa^{-1}m^{2}f\left(  \kappa z\right)
\eqno(37)$$
so that the expectation value of $z$ at the minimum of the potential is  such
that $\kappa\left\langle z\right\rangle =O\left(  1\right)  $ and
$\left\langle g_{2}\right\rangle =O\left(  \kappa^{-1}m^{2}\right)$ . Such
supersymmetry breaking leads to a gravitino mass and low-energy supersymmetric
particles of size $m_{s}=\kappa^{2}\left\langle g_{2}\right\rangle =\kappa
m^{2}.$ If we choose $m\sim10^{10}$ Gev then $m_{s}=O$ is of the order of  $\left(  {\rm {Tev}}
\right)  .$ 
\par
The supersymmetric partner $\chi$ of the field $z$ is the
Goldstino, that is,  the Goldstone fermion arising from the supersymmetry breaking. 
It  gets absorbed by the gravitino making it massive with mass
$m_{{3\over 2}}=\kappa^{-1}\left\langle e^{{1\over 2}{\cal{G}}
}\right\rangle $. Taking  the  unification group $G$ to be $SU(5)$, or $SO(10) $,  the gauge coupling constants unify at
a scale $M_{G}\simeq2\times10^{16}$ Gev and the group breaks to $SU\left(
3\right)  _{C}\times SU\left(  2\right)  _{L}\times U\left(  1\right)  _{Y}.$
It was shown in [28] that, for the superpotential of the form given in 
equation (36) and (37), the gauge hierarchy is preserved for both
$M_{\rm{Pl}}$ and $M_{\rm{G}}.$ For the case $f_{\alpha\beta}\left(
z^{A}\right)  =\delta_{\alpha\beta}$ and a flavor blind K\"{a}hler potential
$\phi\left(  z^{A},z_{A}\right)  $ the effective potential takes the simple
form [27]
$$
V_{\rm{eff}}=\vert{ \partial\tilde{g}\over \partial z^{\alpha}} \vert ^{2}
+m_{0}^{2}z^{\alpha}z_{\alpha}+\left(  B_{0}\tilde{g})
^{\left(  2\right)  }+A_{0}\tilde{g}^{\left(  3\right)  }+\rm{h.c}
\right)  +{1\over 2\kappa^{4}}\left\vert g_{\sigma}G_{\alpha}\left(
T^{\sigma}z\right)  ^{\alpha}\right\vert ^{2}
\eqno(38)$$
where $\tilde{g}$ is the superpotential containing only the quadratic and
cubic functions of the light fields $z^{\alpha},$ i.e. $\tilde{g}\left(
z^{\alpha}\right)  =\tilde{g}^{\left(  2\right)  }\left(  z^{\alpha
}\right)  +\tilde{g}^{\left(  3\right)  }\left(  z^{\alpha}\right)  ,$
$m_{0},$ $A_{0},$ $B_{0}$ are soft breaking parameters of size $m_{s}$ and
$G_{\alpha}=\tilde{g}_{,\alpha}+{\kappa^{2}\over 2}z_{\alpha
}\tilde{g}.$ The most remarkable feature, however, is that the breaking of
supergravity in the hidden sector induces the breaking of $SU\left(  2\right)
_{L}\times U\left(  1\right)  _{Y}$. 
\par
The fact that the super-Higgs mass scale
$m_{s}$ of the soft breaking parameters and the scale of $SU(2)\times U(1)$
breaking are comparable, i.e. both lie in the TeV region, is a natural
consequence of the heavy top quark. The two Higgs doublets have an effective
coupling in the superpotential in the form $\mu_{0}H_{1}H_{2}$ with $\mu_{0}$
is of size $m_{s}.$ Thus one is led to a simple model with five universal
parameters at the GUT scale: $m_{0},$ $m_{{1\over 2}},$  $A_{0},$ $B,$
$\mu_{0}$ where $m_{{1\over 2}}$ is the mass of the gauginos. These
parameters characterize the way the super-Higgs field interacts with the
matter fields. 
\par
While global (rigid) supersymmetry models can accommodate over $134$ soft
breaking parameters, the supergravity models, called variously SUGRA GUT
model, minimal supergravity model, CMSS or mSUGRA allows one to build simple
models that are relatively natural and with a significantly reduced number of
soft terms.  However, experimental 
results over the last few years have  restricted the five parameter space of the models discussed above to 
a rather small volume  and it would seem that one has to consider more complicated models in order to remain consistent 
with experimental results. For a review of the construction of realistic models of supersymmetry see the review of reference [29]. 
\par
In this review we have explained  the formulation of supergravity [3]  that results from gauging the super Poincar\'e group  and  how it contained  an analytic proof of  the invariance of $D=4$, $N=1$ (simple)  supergravity under local supersymmetry transformations. Indeed this method has been used to prove the invariance of essentially all supergravity theories including those in ten and eleven dimensions. These results allowed for the systematic development of supergravity theories and in particular  the discovery of the auxiliary, fields, that is, a formulation of simple supergravity whose transformations closed without  the use of the equations of motion [10,11]. The resulting algebra allowed the construction of the analogue of the tensor calculus for general relativity for simple supergravity[22,23].  This in turn allowed for the construction of the  most general matter coupling to supergravity and as a result  the realistic models making it possible to test low energy supersymmetry experimentally. 

\medskip
{\bf{ Acknowledgments}}
The work of A. H. C is supported in part by the National Science Foundation
Grant No. Phys-1912998 and P.C.W has benefitted from grant numbers ST/P000258/1  and ST/T000759/1  from STFC in the UK. 

%%%%%%%%%%%%

%\medskip
%{\bf References}
%\medskip


\begin{thebibliography}{} 
\bibitem{[1]} D. Freedman, P. van Nieuwenhuizen and S. Ferrara, {\it Progress toward a theory of supergravity }, 
Phys. Rev. {\bf D13}, 3214 (1976); {\it Phys. Rev.} {\bf D14},
912 (1976).
\bibitem{[2]} S. Deser and B. Zumino,  {\it Consistent Supergravity}, {\it Phys. Lett.} {\bf 62B},
\bibitem{[3]} A. Chamseddine and P. West, {\it Supergravity as a gauge theory of supersymmetry}, {\it Nucl. Phys.}  {\bf
B129}, 39 (1977). This paper was received on the 28 September 1976 but was initially rejected for publication. The  revised version,  received  on  10 June 1977, was essentially the same except for two paragraphs explaining in more detail its relation to references [1] and [2]. It was also published as the  Imperial preprint ICTP/75/22, September 1976. 
\bibitem{[4]}  P. West, {\it Introduction to string and branes}, 2012, Cambridge University Press. 
\bibitem{[5]} P. West, {\it Introduction to supersymmetry and
supergravity, } second edition 1990, World Scientific, Singapore. 
\bibitem{[6]} P. Townsend and P. van Nieuwenhuizen, {\it Geometrical interpretation of extended supergravity},  {\it Phys.
Lett.} {\bf B67}, 439 (1977).
\bibitem{[7]}  A. H. Chamseddine, {\it Supersymmetry and higher spin fields},  Ph.D. Thesis, defended September 1976, Imperial College, London University. Chapter 6 of the thesis is stated as collaborative work with P. West. 
The thesis can be found at 
\par
\href{https://drive.google.com/file/d/0B8tITtoqQkxfeHQ3cFJ0WUxIcHM/view?usp=sharing.}{Link to thesis}
\bibitem{[8]}  S. MacDowell and F. Mansouri, {\it Unified geometric theory of gravity and supergravity}, Phys. Rev. Lett. {\bf 38 }(1977) 739. This paper was received on the 9 February 1977. 
\bibitem {[9]} A. H. Chamseddine, {\it Massive supergravity from spontaneously breaking orthosymplectic gauge symmetry}, Ann. Phys. {\bf 113} (1978) 219. 
\bibitem{[10]} K. Stelle and P. West, {\it Minimal auxiliary fields for supergravity}, {\it Phys. Lett.} {\bf B74}, 569. 
330 (1978).   
\bibitem{[11]} S. Ferrara and P. van Nieuwenhuizen, {The auxiliary fields of supergravity}, {\it Phys.
Lett.} {\bf B74}, 333 (1978).
\bibitem{[12]} K. Stelle and P. West, {\it Matter coupling and BRS transformations with auxiliary fields in supergravity}, Nucl. Phys. {\bf B140 }(1978) 285.
\bibitem{[13]} J. Dixon, {The 1.5 order formalism does not generate a BRS transformation for supergravity}, arXiv:2112.11906.
\bibitem{[14]}  P. West, {\it A geometric gravity lagrangian }; Phys. Lett. {\bf 76B}, (1978) 
K.S. Stelle and P. West, {\it de Sitter Gauge Invariance and the Geometry of the
Einstein-Cartan Theory},  J.Phys. {\bf A12}, L205;  (1979); .{\it  Spontaneously Broken de Sitter Symmetry and the Gravitational Holonomy group}, Phys. Rev. {\bf D21} 1466 (1980).
\bibitem{[15]} H. Weyl, {\it Gravitation and the electron}, Proc. N. A. S. {\bf 15 }(1929) 323.
\bibitem{[16]} D. Sciama, {\it The physical structure of general relativity}, Rev. Mod. Phys. {\bf 36 }(1964) 463.
\bibitem{[17]} R. Utiyama, {\it Invariant theoretical interpretation of interaction}, Phys. Rev. {\bf 101 }(1956) 1597.
 \bibitem{[18]} W. B. Kibble, {\it Lorentz invariance and the gravitational field }, J. Math. Phys. {\bf 2 }(1961) 212.
\bibitem{[19]} See for example the collected papers in the reprint volume by F. Hehl , {\it Gauge Theories of Gravitation}, Imperial College Press, London, April 2013. 
\bibitem{[20]} P. Van Nieuwenhuizen, M. Kaku, and P. Townsend, {\it Properties of conformal supergravity}, Phys. Rev. {\bf D17}, (1978) 1501. 
\bibitem{[21]} See for example the review of V Didenko and E.Skvortsov, {\it Elements of Vasiliev theory}, ArXiv 1401.2975. This paper provides  an example of how the contents of reference [3] have entered into the common knowledge although their origin have  been lost. 
\bibitem{[22]} K.S. Stelle and P. West, {Tensor calculus for the vector multiplet coupled to supergravity},  {\it Phys. Lett.} {\bf 77B},
376 (1978);  {\it Relation between vector and scalar multiplets and gauge invariance in supergravity},  Nucl. Phys. {\bf B145} (1978) 175.
\bibitem{[23]} S. Ferrara and P. van Nieuwenhuizen, {\it Tensor calculus for supergravity}, Phys. Lett. {\bf 76B} (1978) 404;  
{\it Structure of supergravity},  Phys. Lett. {\bf 78B}, 573 (1978). 
\bibitem{[24]}  A. H. Chamseddine, R. Arnowitt and P. Nath, {\it Locally
supersymmetric grand unification}, Phys. Rev. Lett. {\bf 49 }(1982) 970.
\bibitem{[25]}  E. Cremmer, S. Ferrara, L. Girardello and A. Van Proeyen,
{\it Coupling supersymmetric Yang-Mills gauge theories to supergravity}, Phys. Lett. {\bf 116}B (1982) 231.
\bibitem{[26]}  P. Nath, R. Arnowitt and A. H. Chamseddine, {\it Applied N=1
Supergravity, }ICTP\ series in Theoretical Physics, Volume 1, (1982), World
Scientific, Singapore. 
\bibitem{[27]} R. Barbieri, S. Ferrara and C. Savoy, {\it Gauge models with
spontaneously broken local supersymmetry, }Phys. Lett. B{\bf  119 }(1982) 343.
\bibitem{[28]}  P. Nath, R. Arnowitt and A. H. Chamseddine, {\it Gauge
heirarchy in supergravity in supergravity Guts}, Nucl. Phys. B{\bf  227
}(1983) 121.
\bibitem{[29]} H. P. Nilles, {\it Supersymmetry, supergravity and particle physics},  Phys. Rep. {\bf 110} (1984) 1. 
\bibitem{[30]} A. Achucarro and P. Townsend, {\it A Chern-Simons Action for Three-Dimensional anti-De
Sitter Supergravity Theories}, Phys.Lett. B180 (1986) 89. E. Witten, {\it (2+1)-Dimensional Gravity as an Exactly Soluble System},  Nucl.Phys. B311
(1988) 46. 
\end{thebibliography}
\end{document}